\documentclass[aps,prl,twocolumn,showpacs,amsmath,floatfix,citeautoscript]{revtex4}

\usepackage{graphicx}% Include figure files
\usepackage{dcolumn}% Align table columns on decimal point
\usepackage{bm}% bold math
\usepackage{color}

\usepackage{SIunits}
\usepackage{comment}
\usepackage{ulem}

\begin{document}

\title{Slow Exciton Spin Relaxation in Single Self-Assembled In$_{1-x}$Ga$_x$As/GaAs Quantum Dots}

\author{Hai Wei}
\affiliation{%
Key Laboratory of Quantum Information, University of Science and Technology of China,
Hefei, 230026, People's Republic of China
}%

\author{Guang-Can Guo}
\affiliation{%
Key Laboratory of Quantum Information, University of Science and Technology of China,
Hefei, 230026, People's Republic of China
}%

\author{Lixin He}
\email{helx@ustc.edu.cn}
\affiliation{%
Key Laboratory of Quantum Information, University of Science and Technology of China,
Hefei, 230026, People's Republic of China
}%

\date{\today}% It is always \today, today,

\begin{abstract}
We calculate the acoustic phonon-assisted exciton spin relaxation 
in single self-assembled In$_{1-x}$Ga$_x$As/GaAs quantum dots using an
atomic empirical pseudopotential method. We show that the transition from
bright to dark exciton states is induced by Coulomb correlation effects. 
The exciton spin relaxation time obtained from sophisticated configuration
interaction calculations is approximately 15--55 $\mu$s 
in pure InAs/GaAs QDs and even longer in alloy dots. 
These results contradict previous theoretical and
experimental results, which suggest very short exciton spin times (a few ns), 
but agree with more recent experiments that suggest 
that excitons have long spin relaxation times ($>$ 1 $\mu$s). 
\end{abstract} 

\maketitle

%\section{Introduction}
%\label{sec:introduction}

Self-assembled quantum dots (QDs) have many attractive features as
fundamental building blocks for quantum information processing. 
However, their short spin lifetime is still a major obstacle  
for such applications.
There have been extensive studies of single electron and hole 
spin relaxation in QDs caused by hyperfine
interaction with nuclear spins~\cite{braun05,kroutvar04} and the spin-phonon interaction due to 
spin-orbit coupling
(SOC)~\cite{cheng04,golovach04,heiss07,gerardot08,khaetskii01,trif09,wei12,warburton13}.
However, exciton spin relaxation has been less commonly studied.

Excitons and biexcitons in QDs have been used to generate single photons~\cite{michler00} or
entangled photon pairs~\cite{stevenson06}.
Bright and dark excitons have also been proposed as possible quantum bits (qubits)~\cite{biolatti00,Imamoglu99}.
The fast nonradiative relaxation of bright excitons limits the maximal single-photon
device emission rate and thus lowers the source efficiency~\cite{strauf07}.
This property also lowers the quality of the single photons and the fidelity of 
the entangled photon pairs generated by biexciton
cascade decay~\cite{reichle08}.
Despite its importance, spin relaxation in excitons is still
not well understood and full of controversy.

Exciton spin relaxation has been measured
by several groups in different types of QDs~\cite{snoke04,johansen10,kurtze11}. The measured
spin relaxation time ranges from 200 ps~\cite{snoke04} 
to 167 ns~\cite{johansen10}. The spin relaxation time calculated from perturbation
theory is approximately 2 ns in In(Ga)As/GaAs QDs at 4 K~\cite{tsitsishvili05},
which seems to be in good agreement with experimental
values~\cite{kurtze11}. All of these studies suggest
fast spin relaxation for excitons.
However, recent direct measurements~\cite{mcfarlane09,poem10} 
of dark exciton lifetimes show that dark excitons
actually have rather long lifetimes ($\sim 1.5\ \mu$s), which serve as a lower
bound for exciton spin relaxation, in sharp contrast to previous results.

To solve the controversy, we calculate the first-order phonon-assisted exciton spin
relaxation in single self-assembled In$_{1-x}$Ga$_x$As/GaAs QDs using an
atomistic empirical pseudopotential method (EPM)~\cite{williamson00}. 
Remarkably, we find that in the Hartree-Fork (HF)
approximation, the transition from a bright to dark state is forbidden,
suggesting that the transition 
is induced by Coulomb correlation effects. 
Sophisticated configuration interaction (CI)~\cite{franceschetti99} calculations suggest that the 
bright-to-dark exciton transition 
is on the order of tens of $\mu s$ in InAs/GaAs QDs, much longer than previous
calculations~\cite{tsitsishvili05} and early experimental values~\cite{snoke04,johansen10,kurtze11}
but supported by more recent measurements~\cite{mcfarlane09}.

%\section{Methods}
%\label{sec:methods}

\begin{figure}
\centering
\includegraphics[width=0.30\textwidth]{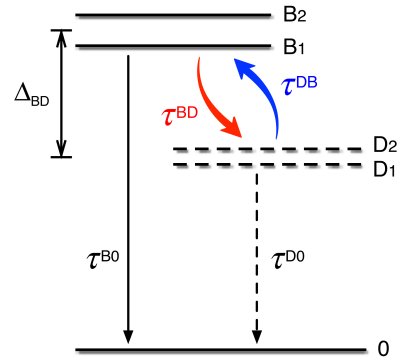}
\caption{(Color online) Schematic depiction of exciton spin relaxation.
$\tau^{BD}$ ($\tau^{DB}$) is the transition time from
  bright (dark) to dark (bright) states.
  $\tau^{B0}$ ($\tau^{D0}$) is the radiative decay time of the bright (dark)
excitons. $\Delta_{BD}$ is the exchange splitting between bright
 and dark excitons. }
\label{fig:cartoon}
\end{figure}

In In$_{1-x}$Ga$_x$As/GaAs QDs, the electron-hole exchange interaction
splits the ground neutral exciton ($X$) states into two optically active (bright)
states with higher energies and two optically inactive (dark) states with lower
energies. Single-dot spectroscopy shows that the typical energy
space between bright and dark states, $\Delta_{BD}$, is approximately 100--300
$\mu$eV~\cite{bayer02}. Because of the asymmetry of the exchange
interaction~\cite{bayer02,bester03,gong11},
the bright (dark) states further split into two sub-levels
$B_1$ and $B_2$ ($D_1$ and $D_2$), as schematically shown in Fig. \ref{fig:cartoon}. 
The energy splitting between $B_1$ and $B_2$, known as fine structure 
splitting (FSS), is usually a few tens of $\mu$eV~\cite{bayer02,gong11}.

In QDs, spin-phonon interaction due to SOC can cause spin flip 
~\cite{cheng04,khaetskii01,golovach04,trif09,wei12}.
In excitons, the spin flip results in
a transition from a bright to dark exciton and the emission of a phonon
or vice versa (see Fig. \ref{fig:cartoon}).
The exciton spin relaxation rate from a bright ($B$) to a dark ($D$) state is
given by the first-order Fermi's Golden Rule:
\begin{eqnarray}\label{equ:rate}
\dfrac{1}{\tau_{\nu}^{BD}} = \dfrac{2\pi}{\hbar} \sum_{\mathbf{q}} 
\left | M_{\nu}^{BD}(\mathbf{q}) \right |^{2} (N_{\nu,\mathbf{q}}+1) \delta(
\Delta_{BD}-\hbar \omega_{\nu,\mathbf{q}})\, ,
\end{eqnarray}
where $N_{\nu,\mathbf{q}}=\left( e^{\hbar \omega_{\nu,\mathbf{q}} / k_B T} -
1\right)^{-1}$ is the Bose-Einstein distribution function for phonons. 
$\hbar \omega_{\nu,\mathbf{q}}$ is the phonon energy, with 
$\omega_{\nu,\mathbf{q}} = c_{\nu} |\mathbf{q}|$,  
where $c_\nu$ is the speed of sound for the $\nu=\text{LA}$ (longitudinal acoustic phonon) and TA
(transverse acoustic phonon) modes.
Because $\Delta_{BD}$ is very small, 
only acoustic phonons are involved in the process. 
The exciton-phonon-coupling matrix element is given by~\cite{takagahara99}
\begin{equation}
M_{\nu}^{BD}(\mathbf{q})
%&=& \langle \Psi_X^D | H_{e-ph} +  H_{h-ph} |\Psi_X^B
%\rangle  \\ \nonumber
=\alpha_{\nu}^e (\mathbf{q}) \langle \Psi_X^D |
e^{i\mathbf{q} \cdot \mathbf{r_e}} | \Psi_X^B \rangle - \alpha_{\nu}^h
(\mathbf{q}) \langle \Psi_X^D | e^{i\mathbf{q} \cdot \mathbf{r_h}} | \Psi_X^B
\rangle  \, ,
\label{equ:Xmatrix}
\end{equation}
where $\Psi_X^B$ ($\Psi_X^D$ are the bright (dark) state wave functions and 
$\alpha_{\nu}^e(\mathbf{q})$ ($\alpha_{\nu}^h (\mathbf{q})$)
is the electron(hole)-phonon-coupling strength.
We have considered three mechanisms in QDs~\cite{cheng04,wei12}, including 
electron(hole)-acoustic-phonon interaction due to
(i) the deformation potential ($\nu=\text{LADP}$), (ii) the piezoelectric field for
the longitudinal modes ($\nu=\text{LAPZ}$), and (iii) the piezoelectric field for
the transverse modes ($\nu=\text{TAPZ}$). Details of
$\alpha_{\nu}^e(\mathbf{q})$ ($\alpha_{\nu}^h (\mathbf{q})$)
and related parameters can be found in Ref.~\onlinecite{wei12}.
The overall spin relaxation time from bright to dark states, $T_1$, is
\begin{eqnarray}
1/T_1 = \sum_{\nu} \sum_{B} \sum_{D} 1/\tau_{\nu}^{BD}\,.
\end{eqnarray}

It is essential to have high-quality exciton wave functions to obtain 
accurate exciton spin relaxation times~\cite{wei12}.
In this work, we use EPM~\cite{williamson00} to calculate
single-particle energy and wave functions. This method has been successfully applied
to study the electronic and optical properties of self-assembled In$_{1-x}$Ga$_{x}$As/GaAs
QDs. We simulate lens-shaped In$_{1-x}$Ga$_{x}$As/GaAs QDs embedded in a cubic
GaAs matrix. We obtain the electron and hole
energy levels and wave functions by solving the 
Schr\"{o}dinger equation via a linear combination of bulk bands (LCBB)
method~\cite{wang99b}, in which the SOC is included in the non-local part of 
the pseudopotentials.
%More details of the calculations can be found in our previous publication~\cite{wei12}.
When single particle wave functions have been obtained, Slater determinants
are built as a basis for excitonic states. The exciton wave functions are obtained via the
CI method~\cite{franceschetti99} by expanding them as a
linear combination of Slater determinants.
The $\alpha\text{-}th$ ($\alpha=D_1,D_2,B_1,B_2$)  exciton wave function is
written as,
\begin{equation}
\Psi_{X}^{\alpha}(\mathbf{r_{e}},\mathbf{r_{h}})
= \sum_{v}^{N_v} \sum_{c}^{N_c} C_{v,c}^{\alpha} \Phi_{v,c}
(\mathbf{r_{e}},\mathbf{r_{h}})\, ,
\label{equ:Xwavefunction}
\end{equation}
where $N_v$ and $N_c$ are the numbers of valence and conduction
states included in the expansion. The coefficients $\{C_{v,c}^{\alpha}\}$ as well
as the  exciton energies are
obtained by diagonalizing the many-particle Hamiltonian in terms of the Slater
determinants basis set $\{\Phi_{v,c}\}$. The exciton energies 
and wave functions are well converged using $N_v$=20 and
$N_c$=12 (including spin) in our calculations.

Once we have obtained exciton wave functions, the
exciton wave function overlap in Eq.~\eqref{equ:Xmatrix} can be calculated as
\begin{equation}
\langle \Psi_{X}^{D} | e^{i\mathbf{q} \cdot \mathbf{r_e}} | \Psi_{X}^{B} \rangle
=  \sum_{v}^{N_v} \sum_{c_i,c_f}^{N_c} (C_{v,c_f}^{D})^{\ast} C_{v,c_i}^{B}
\langle 
\psi^e_{c_f} | e^{i\mathbf{q} \cdot \mathbf{r_e}} | \psi^e_{c_i} \rangle\, ,
\label{equ:Xmatrixe} 
\end{equation}
and 
\begin{equation}
\langle \Psi_{X}^{D} | e^{i\mathbf{q} \cdot \mathbf{r_h}} | \Psi_{X}^{B} \rangle
= \sum_{v_i,v_f}^{N_v} \sum_{c}^{N_c} (C_{v_f,c}^{D})^{\ast}
C_{v_i,c}^{B} 
\langle \psi^h_{v_f} | e^{i\mathbf{q} \cdot \mathbf{r_h}} | \psi^h_{v_i}
\rangle \, ,
\label{equ:Xmatrixh}
\end{equation}
where $\psi^e_{c}$ ($\psi^h_{v}$) is the $c\text{-}th$ ($v\text{-}th$) electron (hole) wave function~\cite{wei12}.

The single particle matrix elements,
$\langle \psi^e_{c_f} | e^{i\mathbf{q} \cdot \mathbf{r_e}} | \psi^e_{c_i}
\rangle$, are calculated in the Bloch basis of bulk InAs at the $\Gamma$ point~\cite{wei12}.
The bright (dark) exciton wave functions are dominated by configurations
in which electron and hole have the opposite (same) pseudo-spin. The mixture of the
configurations, in which electron and hole have the same (opposite) pseudo-spin
because of heavy hole-light hole mixing, is rather small.
Therefore, the matrix elements in Eq.~\eqref{equ:Xmatrixe} %and \eqref{equ:Xmatrixh} 
are expected to be very small because, if electrons $\psi^e_{c_i}$
and $\psi^e_{c_f}$ have the same pseudo spins, ensuring 
that $| \langle \psi^e_{c_f} | e^{i\mathbf{q} \cdot \mathbf{r_e}} | \psi^e_{c_i}
\rangle |$ is large ($\sim$ 1), $| (C_{v,c_f}^{D})^{\ast}C_{v,c_i}^{B} |$ is
small ($<$ 0.01).
However, if electrons $\psi^e_{c_i}$ and $\psi^e_{c_f}$ have opposite pseudo
spins, $|(C_{v,c_f}^{D})^{\ast}C_{v,c_i}^{B}|$ is large ($\sim$ 0.5),
but the single-particle 
wave function overlaps, $|\langle \psi^e_{c_f} | e^{i\mathbf{q} \cdot \mathbf{r_e}} | \psi^e_{c_i}
\rangle|$, must be very small~\cite{wei12}. The same arguments also apply to the holes.

%\section{Results and Discussion}
%\label{sec:results}

%\subsection{Hatree-Fork approximation}
%\label{subsec:HF}

We start with the simplest case, using only the lowest electron and hole
($N_v$=$N_c$=2) states to construct the exciton wave
functions, which is equivalent to the HF approximation.
Surprisingly, we find that the exciton spin relaxation rate is zero in this approximation. 
To understand this result, we examine
Eq.~\eqref{equ:Xmatrixe} and Eq.~\eqref{equ:Xmatrixh} under the HF approximation
in greater detail. 
We first examine the electron part of the exciton wave function overlap,
Eq.~\eqref{equ:Xmatrixe}. Under the HF approximation,
Eq.~\eqref{equ:Xmatrixe} can be written as
\begin{eqnarray}
\label{equ:XmatrixHF}
\langle \Psi_{X}^{D} | e^{i\mathbf{q} \cdot \mathbf{r_e}} | \Psi_{X}^{B} \rangle_{HF}
&=&\xi_{11} \sum_{v=1,2} \left[  (C_{v,1}^{D})^{\ast} C_{v,1}^{B} +
  (C_{v,2}^{D})^{\ast} C_{v,2}^{B} \right]  \\ \nonumber
&+&\xi_{12} \sum_{v=1,2}
(C_{v,1}^{D})^{\ast} C_{v,2}^{B} +\xi_{12}^* \sum_{v=1,2} (C_{v,2}^{D})^{\ast}
C_{v,1}^{B}\, ,
\end{eqnarray}
where $\xi_{11} = \langle \psi^e_{1} | e^{i\mathbf{q} \cdot
  \mathbf{r_e}} | \psi^e_{1} \rangle =
 \langle \psi^e_{2} | e^{i\mathbf{q} \cdot \mathbf{r_e}} |
 \psi^e_{2} \rangle$, and
 $\xi_{12} = \langle \psi^e_{1} | e^{i\mathbf{q} \cdot \mathbf{r_e}}
 | \psi^e_{2} \rangle$. $\psi^e_{1}$ ($\psi^e_{2}$)
are the electron spin up (down) wave functions of the lowest energy level.
Because $\psi^e_1$ and $\psi^e_2$ are Kramers degenerate states that are
related by time reversal symmetry, it is easy to prove that $\xi_{12}$=0. 
Furthermore, the bright ($\Psi_{X}^{B}$) and dark
($\Psi_{X}^{D}$) exciton states are orthogonal:
\begin{eqnarray}
\langle \Psi_{X}^{D} | \Psi_{X}^{B} \rangle_{HF} = \sum_{v=1,2} 
\left[ (C_{v,1}^{D})^{\ast} C_{v,1}^{B} + (C_{v,2}^{D})^{\ast} C_{v,2}^{B}
  \right] = 0\, .
\label{equ:HForthogonalization}
\end{eqnarray}
By substituting $\xi_{12}$ and Eq.~\eqref{equ:HForthogonalization}
into Eq.~\eqref{equ:XmatrixHF}, we have the electron part of the exciton wave function
overlap, $\langle \Psi_{X}^{D} | e^{i\mathbf{q} \cdot \mathbf{r_e}} |
\Psi_{X}^{B} \rangle_{HF}$=0. For the same reason, the hole part of the exciton
wave function overlap is $\langle \Psi_{X}^{D} |
e^{i\mathbf{q} \cdot \mathbf{r_h}} | \Psi_{X}^{B} \rangle_{HF}$=0.
Therefore, the exciton-phonon interaction matrix element, $M_{\nu}^{BD}(\mathbf{q})$=
0,  meaning the exciton spin relaxation rate equals zero under the HF
approximation. Because $\xi_{12} \sim$ 1 is very large, a small
un-orthogonality of the exciton wave functions 
may cause huge errors in the calculated spin relaxation time.
As it is prohibited in the HF approximation, 
exciton spin relaxation is induced by Coulomb correlation effects;
therefore, electron/hole correlation effects (via CI calculations) 
must be included to obtain the correct
relaxation time.

%\subsection{Full Configuration Interaction}
%\label{subsec:CI}

The CI-calculated exciton relaxation times of  pure InAs/GaAs QDs are
approximately 15--55 $\mu$s. The spin relaxation times of alloy QDs are even longer. 
The exciton spin relaxation times are determined by three factors: 
(i) $\Delta_{BD}$, which determines phonon
momentum $\mathbf{q}$ (smaller $\Delta_{BD}$ leads to a
longer spin relaxation time because of lower phonon density), 
(ii) the electron and hole single-particle wave functions, which determine
single-particle relaxation time, and (iii)
the exciton CI coefficients $\{C_{v,c}\}$. 
All three factors are strongly affected by
the geometry and chemical composition of the QDs. 

\begin{figure}
\centering
\includegraphics[width=0.40\textwidth]{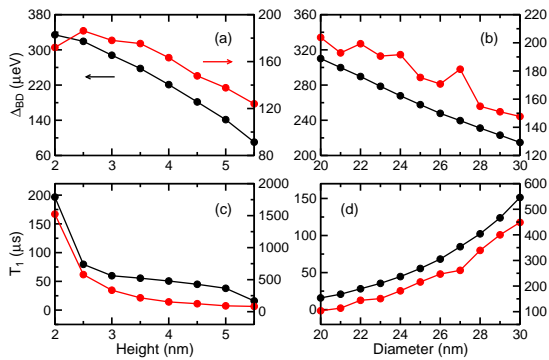}
\caption{ (Color online) 
The upper panels show the exchange energy splitting between bright and dark
states  $\Delta_{BD}$ as functions of
(a) dot height with dot diameter fixed at 25 nm and (b) dot diameter with dot height
fixed at 3.5 nm. Corresponding exciton spin relaxation times are shown in (c) and (d).
Black lines are the results of pure InAs/GaAs QDs, 
whereas red lines are the results of In$_{0.7}$Ga$_{0.3}$As/GaAs QDs.
}
\label{fig:geom}
\end{figure}

We calculate the exciton relaxation time of lens-shaped pure InAs/GaAs QDs as 
a function of base diameter
(height) while keeping height (base diameter) constant. 
The exciton spin relaxation times as well as $\Delta_{BD}$ are given in
Fig.~\ref{fig:geom}  as black lines. 
The calculated $\Delta_{BD}$ distributes mostly between 100--300 $\mu$eV, 
which is in good agreement with experimental values~\cite{bayer02}.
For pure dots, when dot height increases from 2.0 to 5.5 nm, 
$\Delta_{BD}$ decreases from 330 to 90 $\mu$eV [Fig.~\ref{fig:geom}(a)].
We find that the exciton spin relaxation time is dominated by the hole spin flip.
The decrease of $\Delta_{BD}$ tends to slow the spin relaxation time. 
At the same time, the hole spin flip time drops quickly with
increasing dot height~\cite{wei12}.
These two factors compete with each other, and
the overall effect is that the spin flip time decreases first when dot height
changes from 2.0 to 3.0 nm, reaching a relatively constant value as dot height further
increases [Fig.~\ref{fig:geom}(c)]. On the contrary, as the base diameter increases
from 20 to 30 nm, $\Delta_{BD}$ decreases from 310 to
220 $\mu$eV [Fig.~\ref{fig:geom}(b)], whereas increasing dot diameter also
slows hole spin relaxation~\cite{wei12}, and the exciton spin
relaxation time increases [Fig.~\ref{fig:geom}(b)].

\begin{figure}
\centering
\includegraphics[width=0.30\textwidth]{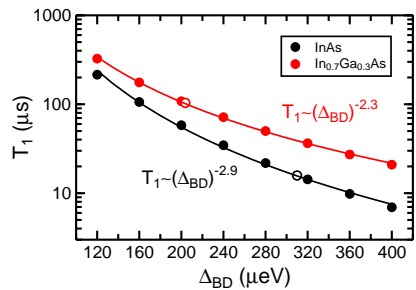}
\caption{
  (Color online) The change of exciton spin relaxation times $T_1$ as we
  artificially vary $\Delta_{BD}$ in lens-shaped In$_{1-x}$Ga$_x$As/GaAs QDs with
  base diameter $b$=20 nm and height $h$=3.5 nm. The open circles are
  results obtained by EPM, and the solid lines are fitted using $T_1\sim\Delta_{BD}^{-\gamma}$.
}
 \label{fig:xrate1_Ebd}
\end{figure}

We also calculate the exciton spin relaxation time
of lens-shaped alloy In$_{0.7}$Ga$_{0.3}$As/GaAs QDs. The results are shown
in Fig.~\ref{fig:geom} as red lines. These results are
similar to those of pure QDs.
Generally, the exciton spin relaxation time of alloy dots due to first-order
spin-phonon coupling is much longer than that of the pure dots because alloy dots usually
have smaller $\Delta_{BD}$. In fact, exciton spin relaxation time is very
sensitive to $\Delta_{BD}$, as demonstrated in Fig. \ref{fig:xrate1_Ebd}. 
We compare the spin relaxation times of two QDs. 
One QD is a lens-shaped InAs/GaAs dot with diameter=20 nm and 
height=3.5 nm. The other QD is an alloy dot with the same geometry but with
Ga composition of $x$=0.3. As we artificially change $\Delta_{BD}$, 
the spin relaxation times increase 
dramatically with decreasing $\Delta_{BD}$, as $T_1\sim\Delta_{BD}^{-\gamma}$,
where $\gamma=2.9$ for the pure dot and $\gamma=2.3$ for the alloy dot from the numerical fit.

\begin{figure}
\centering
\includegraphics[width=0.30\textwidth]{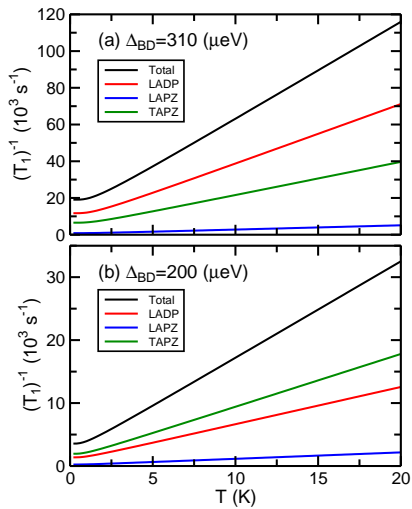}
\caption{
  (Color online) Exciton spin relaxation rates of different mechanisms as a function of
  temperature in lens-shaped InAs/GaAs QDs with base diameter $b$=20 nm
  and height $h$=3.5 nm. The red, blue and green lines denote the LADP,
  LAPZ, and TAPZ contributions, respectively. 
  The black line is the total relaxation rate. (a) $\Delta_{BD}=310$ $\mu$eV, 
  and (b) $\Delta_{BD}=200$ $\mu$eV.
}
\label{fig:mechanism}
\end{figure}

The exciton dark-bright splitting $\Delta_{BD}$ also determines 
which mechanism is dominant for the spin flip. 
Figure~\ref{fig:mechanism} depicts the contributions of the three exciton-phonon interaction
mechanisms to the total exciton spin relaxation rate as a function of
temperature. We take a lens-shaped InAs/GaAs QDs with base diameter $b$=20 nm and
height $h$=3.5 nm as an example. 
The red, blue and green line denote the LADP,
LAPZ, and TAPZ contributions, respectively, whereas
the black line is the total spin relaxation rate. In the experimental
temperature range ($>$4 K), the spin relaxation rates from 
all mechanisms increase linearly with temperature, which is the
signature of the first-order phonon processes.
If we use $\Delta_{BD}$=310 $\mu$eV, which is given by the EPM
calculation, then the LADP mechanism contributes the most to the total relaxation rate, as
shown in Fig.~\ref{fig:mechanism}(a). However, if we (artificially) use a
smaller $\Delta_{BD}$=200 $\mu$eV, then the TAPZ mechanism contributes the most,
as shown in Fig.~\ref{fig:mechanism}(b), because the exciton-phonon coupling strength, 
$\alpha_{\textrm LADP} \propto |\mathbf{q}|$, whereas
$\alpha_{\textrm TADP} \propto 1/|{\bf q}|$. 
Therefore a smaller $\Delta_{BD}$ makes the TAPZ mechanism dominant.

%{\bf Discussion}

Two exciton spin relaxation mechanisms
have been discussed in the literature, namely exchange
interaction~\cite{tsitsishvili03b} 
and SOC~\cite{tsitsishvili05}.  In QDs smaller than the exciton Bohr radius, 
it has been suggested that 
the exchange interaction is most significant in exciton spin relaxation, 
whereas the SOC mechanism dominates in the larger QDs studied here. 
The calculated spin-flip time is approximately 2 ns in the In(Ga)As/GaAs QDs
at 4 K~\cite{tsitsishvili05},
which is a few orders of magnitude faster than that obtained 
in the present work. However, in the previous calculations, the SOC were treated perturbatively.
Cheng \textit{et al.} have shown that, in the
single-particle case, perturbation theory greatly overestimates the
spin relaxation rate~\cite{cheng04}.
The fast relaxation from perturbation theory is
due to the failure of perturbation theory to 
ensure the orthogonality of both
single-particle and many-particle wave functions. 
The (un-orthogonal wave functions) approximation may not be
a serious problem in other calculations;
however, it is crucial in the present calculation as well as in the single
particle case. Because $\xi_{11} \gg
\xi_{12}$, a small error in the  orthogonality of the wave 
functions causes large errors as discussed for the HF approximation above.

The exciton spin relaxation time has been measured by several groups for
different QDs.
Kurtze \textit{et al.}  found that the spin-flip time
is approximately 20 ns at 5 K and 1 ns at 110 K in In(Ga)As/GaAs QDs~\cite{kurtze11}.
Snoke \textit{et al.} found that the dark-to-bright exciton
transition time is 200 ps at T $\sim$ 10 K in InP QDs~\cite{snoke04}.
Johansen \textit{et al.} found that the relaxation time is approximately 77--167 ns in
In(Ga)As/GaAs QDs~\cite{johansen10}. These experimental values
seem to be in good agreement with previous theoretical results
~\cite{tsitsishvili05}, all suggesting that the spin relaxation in excitons is
very fast.
However, in these experiments, spin relaxation times were
extracted from the bright exciton decay time, in which the exciton
radiative decay is much faster than the spin relaxation. Therefore, there
might be very large errors in estimating the spin relaxation time using bright
exciton dynamics. A more accurate
method for estimation of the exciton spin relaxation time is to measure the dark exciton
lifetime, in which the radiative lifetime is extremely long. Indeed, direct
manipulation of dark exciton has recently become possible~\cite{mcfarlane09,poem10}.
The measured dark exciton lifetime exceeds 1.5 $\mu$s at 5 K~\cite{mcfarlane09},
which is the lower bound for the dark-to-bright exciton transition 
(Note that at this temperature, $\tau^{BD} / \tau^{DB} \approx$ 1/2),
ruling out the fast spin relaxation in the exciton. 
This result is supported by the present calculations.

We would like to note that given the very long exciton
relaxation time calculated here, the spin relaxation time through a first-order
spin-phonon interaction may not be the dominant mechanism for
exciton spin relaxation. The roles of other mechanisms need to be further
clarified, including hyperfine and second-order spin-phonon
interactions. Nevertheless, the exciton spin relaxation time should be much
longer than previously reported, which favors quantum information processing.  

%\section{Conclusion}
%\label{sec:conclusion}

To conclude, we present an atomistic pseudopotential calculation of the
acoustic phonon-assisted exciton spin relaxation from bright to dark
exciton in single self-assembled In$_{1-x}$Ga$_x$As/GaAs QDs. We show
that the exciton spin relaxation rate is induced by Coulomb correlation effects.
The spin relaxation time calculated is 15--55 $\mu$s in pure InAs/GaAs
QDs and even longer in alloy dots.
The slow spin relaxation in excitons contradicts previous theoretical and
experimental results, which claim a very short exciton spin lifetime,
but agrees with more recent experiments.

LH acknowledges support from the Chinese National
Fundamental Research Program 2011CB921200
and National Natural Science Funds for Distinguished Young Scholars.

%\bibliographystyle{apsrev} 
%\bibliography{bib/MyDotBiblio2,bib/DotBiblio,bib/footnote}

\end{document}